\documentclass[final,3p,times]{elsarticle}

\usepackage{amssymb}
\usepackage{booktabs}

\usepackage{amssymb}
\usepackage[]{amsmath}
\usepackage{graphics}
\usepackage{amssymb}
\usepackage[]{amsmath}
\usepackage{amsthm}
\usepackage{setspace}
\usepackage{epsfig}
\usepackage{subfigure}
\usepackage{booktabs}

\biboptions{numbers,sort&compress}
\usepackage[dvipdfm,colorlinks,linkcolor=red,anchorcolor=blue,citecolor=green]{hyperref}

\usepackage{amsmath}
\usepackage{times}
\usepackage{anysize}
\marginsize{2.5cm}{2.5cm}{1cm}{2cm}

\selectfont

 \journal {}

\begin{document}

\begin{frontmatter}




\begin{center}
{\LARGE\bf  Explicit solutions , conservation laws of the extended (2+1)-dimensional Jaulent-Miodek equation}
\footnotetext{\hspace{-.55cm}{}
\\$^*$ Corresponding author Email: liyu910118@sina.com}
\end{center}

\begin{description}
\item[{}\mbox{}\quad]\ \

\begin{center}
Yu Li$^{*}$, Xiqiang Liu and Xiangpeng Xin
\end{center}
\begin{center}
{\emph{School of Mathematical Sciences, Liaocheng University, Liaocheng 252059, People's Republic of China}}\\[5pt]

\end{center}

\end{description}


%

\begin{abstract}
By applying the direct symmetry method, the symmetry reductions and some new group invariant solutions were obtained, We have derived some exact solutions by using the relationship between the new solutions and the old ones, which include Weierstrass periodic solutions, elliptic periodic solutions, triangular function solutions and so on. Also, in order to reflect the characteristics and properties of this solutions, we give figures of some solutions. In addition, we give the conservation laws of the extended (2+1)-dimensional Jaulent-Miodek equation. At last, we draw conclusions and discuss it.

\end{abstract}

\end{frontmatter}


\section{Introduction}

The nonlinear evolution equations(NLEEs) are encountered in a variety of scientific fields, such as chemistry, plasma physics, engineering , etc. In order to solve the problems of the nonlinear evolution equations, so many effective analytical and semianalytical approaches have been introduced, such as Lie point symmetry method\cite{1,2,3}, Exp-function method\cite{4}, tanh-function method\cite{5,6}, Jacobi-elliptic method\cite{7,8}, F-expansion method\cite{9}, Painleve analysis and the Hirota bilinear method\cite{10,11}. Especially, to find the Lie point symmetry and reduction of a given partial differential equation, the classical Lie group approach, the non-classical Lie group approach,the direct symmetry method have been suggested.

In this paper, we consider the extended (2+1)-dimensional Jaulent-Miodek equation. The Jaulent-Miodek equation refers to many branches of physics, such as condensed matter physics, fluid dynamics and optics. The relationship between Jaulent-Miodek equation and Euler-Darboux equation is found by Matsuno\cite{12}. In recent years, many scientists have used many methods to obtain the exact solutions of Jaulent-Miodek equation, and have done a lot of work closely related to the spectral problem of Jaulent and Miodek\cite{13,14}. Fan\cite{15}gives the Jaulent-Miodek equation with some traveling wave solutions, and Feng\cite{16} gives the display solutions to the problem. For example, Exp-function method\cite{4} is used to get the equation with periodic solution, tanh-coth method obtained the equation of soliton solutions, sign calculation equation is shaped as a cusp like singular soliton solutions, F - function equation of the traveling wave solutions are obtained\cite{17,18,19,20}.

The generaliged (2+1)-dimensional Jaulent-Miodek equation reads
\begin{equation}\label{ch-0}
u_{xt}  - \alpha u_{xxxx}  + \beta  u_x^2 u_{xx}  - u_{xx} u_y + \gamma u_x u_{xy}  + \lambda u_{yy} + \delta u_{xx} = 0,
\end{equation}
$\alpha,\beta,\gamma,\lambda$ and $\delta$ are real constants. The (2+1)-dimensional Jaulent-Miodek equation is generated by the (1+1)-dimensional Jaulent-Miodek hierarchy were developed\cite{21}. When $\delta=0$, a variety of solutions of the equation are obtained, please refer to the literature \cite{22}.

By applying the direct symmetry method, we get the symmetry reductions, group invariant solutions and some new exact solutions of  Eq.(\ref{ch-0}), which include Weierstrass periodic solutions, Elliptic periodic solutions, Triangular function solutions and so on. This paper is organized in the following sections. In Sec.\ref{local}. the Lie point symmetry groups and symmetry are obtain. And the brackets and the single parameter group are obtained, also the invariant solution of the group is obtained by the single parameter group. In Sec.\ref{similarity reduction}. we find the reduction and new solutions of the Eq.(\ref{ch-0}). In Sec.\ref{Conservation laws}. we give the conservation laws of Eq.(\ref{ch-0}). At last, we draw conclusions and discuss it.

\section{Symmetry groups of the extended (2+1)-dimensional Jaulent-Miodek equation}\label{local}

By applying the classical Lie symmetry method we consider the one-parameter
 Lie group of infinitesimal transformations in $(x,y,t,u)$ of Eq.(\ref{ch-0}) given by
 \begin{center}
$\begin{array}{l}
 x^\ast=x + \epsilon \xi \left( {x,y,t,u} \right) + {\rm O}\left( {\epsilon ^2 } \right),\\
 y^\ast=y + \epsilon \eta \left( {x,y,t,u} \right) + {\rm O}\left( {\epsilon ^2 } \right),\\
 t^\ast=t + \epsilon \tau \left( {x,y,t,u} \right) + {\rm O}\left( {\epsilon ^2 } \right),\\
 u^\ast=u + \epsilon \phi \left( {x,y,t,u} \right) + {\rm O}\left( {\epsilon ^2 } \right),\\
 \end{array}$
\end{center}
where $\epsilon$ is a small parameter. Thus, the vector field with the above group of transformations can be expressed as follows
 \begin{equation}\label{ch-1}
V  = \xi(x,y,t,u) \frac{\partial }{{\partial x}} + \eta(x,y,t,u) \frac{\partial }{{\partial y}} + \tau(x,y,t,u) \frac{\partial }{{\partial t }} + \phi(x,y,t,u) \frac{\partial }{{\partial u }},
\end{equation}
where the coefficient functions $\xi,\eta,\tau$ and $\phi$ are to be determined later.

If the vector field Eq.(\ref{ch-0}) generates a symmetry of Eq.(\ref{ch-1}) , then must satisfy Lie's symmetry condition
\begin{equation}\label{ch-2}
 P r^{(4)} V (\Delta)|_{\Delta = 0} = 0,
\end{equation}
where $\Delta = u_{xt}  - \alpha u_{xxxx}  + \beta  u_x^2 u_{xx}  - u_{xx} u_y + \gamma u_x u_{xy}  + \lambda u_{yy} + \delta u_{xx}$. Applying the fourth prolongation $P r^{(4)} V$ to Eq.(\ref{ch-0}), we find the following system of symmetry equations, and the invariant condition reads as
\begin{equation}\label{ch-3}
2\beta \phi^x u_x u_{xx}^2 + \gamma \phi^x u_{xy} + \phi^{xt} - \phi^{xx} u_y + \beta \phi^{xx} u_x^2 - \phi^y u_{xx} + \gamma \phi^{xy} u_x + \lambda \phi^{yy} + \delta \phi^{xx} - \alpha \phi^{xxxx} = 0,
\end{equation}
where $\phi^x,\phi^{xx},\phi^{xt},\phi^{xy},\phi^y,\phi^{yy},\phi^{xxxx}$ are given explicitly in terms of $\xi,\eta,\tau,\phi$ and the derivatives of $\phi$. Furthermore, one can get

\begin{equation}\label{eq4}
\begin{array}{lll}
 \phi^x=D_x(\phi - u_x \xi - u_t \tau - u_y \eta) + \xi u_{xx} + \eta u_{xy} + \tau u_{xt},\\
 \phi^y=D_y(\phi - u_x \xi - u_t \tau - u_y \eta) + \xi u_{xy} + \eta u_{yy} + \tau u_{yt},\\
 \phi^{xx}=D_{xx}(\phi - u_x \xi - u_t \tau - u_y \eta) + \xi u_{xxx} + \eta u_{xxy} + \tau u_{xxt},\\
 \phi^{xt}=D_{xt}(\phi - u_x \xi - u_t \tau - u_y \eta) + \xi u_{xxt} + \eta u_{xyt} + \tau u_{xtt},\\
 \phi^{xy}=D_{xy}(\phi - u_x \xi - u_t \tau - u_y \eta) + \xi u_{xxy} + \eta u_{xyy} + \tau u_{xyt},\\
 \phi^{yy}=D_{yy}(\phi - u_x \xi - u_t \tau - u_y \eta) + \xi u_{xyy} + \eta u_{yyy} + \tau u_{yyt},\\
 \phi^{xxxx}=D_{xxxx}(\phi - u_x \xi - u_t \tau - u_y \eta) + \xi u_{xxxxx} + \eta u_{xxxxy} + \tau u_{xxxxt},
 \end{array}
\end{equation}
where $D_i$ denotes the total derivative operator and is defined as
$D_i = \frac{\partial}{\partial x_i} + u_i \frac{\partial}{\partial u} + u_{ij} \frac{\partial}{\partial x_j} + \cdots$, and $(x_1,x_2,x_3) = (t,x,y)$.

By Lie symmetry analysis, setting the coefficients of the polynomial to zero, yields many differential equations about the function $\xi,\eta,\tau$ and $\phi$, and solve the corresponding determined equations, one can get
\begin{equation}\label{ch-4}
\begin{array}{l}
 \xi = \frac{c_1}{3} x - F_1(t) + \frac{2c_1 \delta t}{3} + \frac{2\beta c_3 y}{-4\beta \lambda + \gamma^2 - \gamma} + c_5, \\
 \eta = \frac{2c_1}{3} y +c_3 t +c_4 , \\
 \tau = c_1 t + c_2 , \\
 \phi = \frac{c_3 (\gamma-1) x}{-4\beta \lambda + \gamma^2 - \gamma} + F'_1(t) y +  F_2(t) ,\\
 \end{array}
\end{equation}
where $c_i(i=1,2,3,4,5)$ are arbitrary constants, $F_1(t)$ and $F_2(t)$ are arbitrary functions.  Therefore, we obtain the following symmetries of Eq.(\ref{ch-0}),
\begin{equation}\label{ch-5}
\begin{array}{l}
 \sigma  = \left( {\frac{{c_1 }}{3}x - F_1 \left( t \right) + \frac{{2c_1 \delta t}}{3} + \frac{{2\beta c_3 y}}{{ - 4\beta \lambda  + \gamma ^2  - \gamma }} + c_5 } \right)u_x  + \left( {\frac{{2c_1 }}{3}y + c_3 t + c_4 } \right)u_y + \left( {c_1 t + c_2 } \right)u_t  \\  ~~~~~~~~- \left( {\frac{{c_3 \left( {\gamma  - 1} \right)x}}{{ - 4\beta \lambda  + \gamma ^2  - \gamma }} + F_1 ^\prime  \left( t \right)y + F_2 \left( t \right)} \right) .\\
 \end{array}
\end{equation}

From the above results, equivalent vector expression of the above symmetry is
\begin{equation}\label{ch-6}
\begin{array}{l}
 V = \left( {\frac{{c_1 }}{3}x - F_1 \left( t \right) + \frac{{2c_1 \delta t}}{3} + \frac{{2\beta c_3 y}}{{ - 4\beta \lambda  + \gamma ^2  - \gamma }} + c_5 } \right)\frac{\partial }{{\partial x}} + \left( {c_1 t + c_2 } \right)\frac{\partial }{{\partial t}} + \left( {\frac{{2c_1 }}{3}y + c_3 t + c_4 } \right)\frac{\partial }{{\partial y}} \\ ~~~~~~~~+ \left( {\frac{{c_3 \left( {\gamma  - 1} \right)x}}{{ - 4\beta \lambda  + \gamma ^2  - \gamma }} + F_1 ^\prime  \left( t \right)y + F_2 \left( t \right)} \right)\frac{\partial }{{\partial u}} .\\
\end{array}
\end{equation}

Setting $F_1(t)=c_6 t + c_7$, $F_2(t)=c_8 t^2+c_9 t + c_{10}$, where $c_i(i=6,7,8,9,10)$ are arbitrary constants. From Eq.(\ref{ch-6}), the corresponding ten-dimensional Lie algebra $= \{V_1,V_2,V_3,V_4,V_5,V_6,V_7,V_8,V_9,V_{10}\}$ is reached, where
\begin{equation}\label{ch-7}
\begin{array}{l}
 V_1  = (\frac{1}{3} x + \frac{2 \delta t}{3}) \frac{\partial}{\partial x} + \frac{2}{3} y \frac{\partial}{\partial y} + t \frac{\partial}{\partial t}, V_2 = \frac{\partial}{\partial t}, \\
 V_3  = \frac{2\beta y}{-4\beta \lambda + \gamma^2 - \gamma} \frac{\partial}{\partial x} + t \frac{\partial}{\partial y} + \frac{(\gamma-1) x}{-4\beta \lambda + \gamma^2 - \gamma} \frac{\partial}{\partial u},\\
 V_4  = \frac{\partial}{\partial y},~~~~V_5  = \frac{\partial}{\partial x},~~~~V_6  = -t \frac{\partial}{\partial x} + y \frac{\partial}{\partial u}, \\
 V_7  = -\frac{\partial}{\partial x},~~~~V_8  = t^2 \frac{\partial}{\partial u},~~~~V_9  = t \frac{\partial}{\partial u}, \\  V_{10}  =  \frac{\partial}{\partial u},
 \end{array}
\end{equation}

 The one-parameter group generated by $v_i(i=1,2,\cdots,10)$ is given as
 \begin{equation}\label{ch-8}
\begin{array}{l}
 g_1 :\left( {t\delta e^\varepsilon   + \left( { - t\delta  + x} \right)e^{\frac{\varepsilon }{3}} ,ye^{\frac{2}{3}\varepsilon } ,te^\varepsilon  ,u} \right),
g_2 :\left( {x,y,t + \varepsilon ,u} \right),  \\
g_3 :\left( {\int_0^\varepsilon  {\frac{{2\beta \left( {z_1 t + y} \right)}}{{ - 4\beta \lambda  + \gamma ^2  - \gamma }}} dz_1  + x,y + t\varepsilon ,t,\int_0^\varepsilon  {\frac{{\left( {\gamma  - 1} \right)\left( {\int_0^{z_1 } {\frac{{2\beta \left( {z_1 t + y} \right)}}{{ - 4\beta \lambda  + \gamma ^2  - \gamma }}dz_1  + x} } \right)}}{{ - 4\beta \lambda  + \gamma ^2  - \gamma }}dz_1  + u} } \right),  \\
g_4 :\left( {x,y + \varepsilon ,t,u} \right) , g_5 :\left( {\varepsilon  + x,y,t,u} \right) , g_6 :\left( {x - t\varepsilon ,y,t,y\varepsilon  + u} \right),  \\
g_7 :\left( { - \varepsilon  + x,y,t,u} \right) ,  g_8 :\left( {x,y,t,t^2 \varepsilon  + u} \right) ,  g_9 :\left( {x,y,t,t \varepsilon  + u} \right) , \\ g_{10} :\left( {x,y,t, \varepsilon  + u} \right).
 \end{array}
\end{equation}

From the above groups $g_i(i=1,2,\cdots,10)$ , one can get the following solutions of Eq.(\ref{ch-0})
\begin{equation}\label{ch-9}
\begin{array}{l}
u_1  = f\left( { - \frac{{\delta te^\varepsilon   - \delta te^{\frac{\varepsilon }{3}}  - xe^\varepsilon  }}{{e^{^{\frac{\varepsilon }{3}} } e^\varepsilon  }},ye^{ - \frac{2}{3}\varepsilon } ,te^{ - \varepsilon } } \right) , u_2  = f\left( {x,y,t - \varepsilon } \right) , \\
u_3  = f\left( {x - \int_0^\varepsilon  {\frac{{2\beta \left( {z_1 t - t\varepsilon  + y} \right)}}{{ - 4\beta \lambda  + \gamma ^2  - \gamma }}dz_1 } ,y - t\varepsilon ,t} \right) + \int_0^\varepsilon  {\frac{{\left( {\gamma  - 1} \right)\left( {\int_0^{z_1 } {\frac{{2\beta \left( {z_1 t + y} \right)}}{{ - 4\beta \lambda  + \gamma ^2  - \gamma }}dz_1  + x} } \right)}}{{ - 4\beta \lambda  + \gamma ^2  - \gamma }}dz_1 } , \\
u_4  = f\left( {x,y - \varepsilon ,t} \right) , u_5  = f\left( {x - \varepsilon ,y,t} \right) , u_6  = f\left( {x + t\varepsilon ,y,t} \right) + y\varepsilon , \\  u_7  = f\left( {x + \varepsilon ,y,t} \right) , u_8  = f\left( {x,y,t} \right) + t^2 \varepsilon ,u_9  = f\left( {x,y,t} \right) + t \varepsilon ,\\u_{10}  = f\left( {x,y,t} \right) +  \varepsilon ,
 \end{array}
\end{equation}
where $\epsilon$ is a parameter , $f$ is an arbitrary known solution of Eq.(\ref{ch-0}).

\section{Similarity reductions and new exact solutions of the extended (2+1)-dimensional Jaulent-Miodek equation}\label{similarity reduction}

To search for more similarity reductions of (\ref{ch-0}), one should consider Lie point symmetries of (\ref{ch-0}).In this section, we will discuss how to construct similarity reductions and solutions of Eq.(\ref{ch-0}). The characteristic equations have been written from Eq.(\ref{ch-5}) as follows
\begin{equation}\label{ch-10}
\frac{{dx}}{{\frac{{c_1 }}{3}x - F_1 \left( t \right) + \frac{{2c_1 \delta t}}{3} + \frac{{2\beta c_3 y}}{{ - 4\beta \lambda  + \gamma ^2  - \gamma }} + c_5 }} = \frac{{dy}}{{\frac{{2c_1 }}{3}y + c_3 t + c_4 }} = \frac{{dt}}{{c_1 t + c_2 }} = \frac{{du}}{{\frac{{c_3 \left( {\gamma  - 1} \right)x}}{{ - 4\beta \lambda  + \gamma ^2  - \gamma }} + F_1 ^\prime  \left( t \right)y + F_2 \left( t \right)}}.
\end{equation}

Here we discuss the following Cases.

\textbf{Case 1}~~~~Letting $c_2  = c_4  = 1,c_1  = c_3  = c_5  = 0,F_1 \left( t \right) = 0,F_2 \left( t \right) \ne 0$ , solving the corresponding characteristic equations, we have the following invariants $\xi,\eta$ and the function $u$
\begin{equation}\label{ch-11}
\begin{array}{l}
\xi  = x,\eta  = t - y ,
~~~~~~u = \int {F_2 \left( t \right)dt}  + f\left( {\xi ,\eta } \right),  \\
 \end{array}
\end{equation}

Substituting Eq.(\ref{ch-11}) into Eq.(\ref{ch-0}) , we have
\begin{equation}\label{ch-12}
f_{\xi \eta }  - \alpha f_{\xi \xi \xi \xi }  + \beta f_\xi  ^2 f_{\xi \xi }  + f_{\xi \xi } f_\eta   - \gamma f_\xi  f_{\xi \eta }  + \lambda f_{\eta \eta }  + \delta f_{\xi \xi }  = 0 ,
\end{equation}

In order to get the exact solution of Eq.(\ref{ch-12}), we look for the Lie point transformation group of Eq.(\ref{ch-12}), and the corresponding vector field of Eq.(\ref{ch-12}) as follows
\begin{equation}\label{ch-13}
\widetilde{V_1} = \left( {\frac{a}{2}\xi  + \frac{{a\eta \beta }}{{4\beta \lambda  - \gamma ^2  + \gamma }} + d} \right)\frac{\partial }{{\partial \xi }} + \left( {a\eta  + b} \right)\frac{\partial }{{\partial \eta }} + \left( { - \frac{{a\left( {\gamma  - 1} \right)\xi }}{{8\beta \lambda  - 2\gamma ^2  + 2\gamma }} - \frac{{a\left( { - 4\beta \delta \lambda  + \delta \gamma ^2  + \beta  - \delta \gamma } \right)\eta }}{{ - 4\beta \lambda  + \gamma ^2  - \gamma }} + c} \right)\frac{\partial }{{\partial f}} ,
\end{equation}
where $a,b,c$ and $d$ are arbitrary constants.

Furthermore, we can get the characteristic equation of Eq.(\ref{ch-12}),
\begin{equation}\label{ch-14}
\frac{{d\xi }}{{\frac{a}{2}\xi  + \frac{{a\eta \beta }}{{4\beta \lambda  - \gamma ^2  + \gamma }} + d}} = \frac{{d\eta }}{{a\eta  + b}} = \frac{{df}}{{ - \frac{{a\left( {\gamma  - 1} \right)\xi }}{{8\beta \lambda  - 2\gamma ^2  + 2\gamma }} - \frac{{a\left( { - 4\beta \delta \lambda  + \delta \gamma ^2  + \beta  - \delta \gamma } \right)\eta }}{{ - 4\beta \lambda  + \gamma ^2  - \gamma }} + c}} ,
\end{equation}
where $a,b,c$ and $d$ are arbitrary constants.

\textbf{Case 1.1}~~~~$a = 0,c = 0,b \ne 0,d \ne 0$ \\

In this case, we can get the following solution of Eq.(\ref{ch-12})
\begin{equation}\label{ch-15}
\theta  = b\xi  - d\eta , ~~~~~~~~f = g\left( \theta  \right) , \\
\end{equation}
and $g(\theta)$ need satisfy
\begin{equation}\label{ch-16}
\left( {\lambda d^2  + \delta b^2  - bd} \right)g'' - \alpha b^4 g^{\left( 4 \right)}  + \beta b^4 g'^2 g'' + \left( {\gamma b^2 d - b^2 d} \right)g'g'' = 0 ,
\end{equation}

Integral to Eq.(\ref{ch-16}) , yields
\begin{equation}\label{ch-17}
\left( {\lambda d^2  + \delta b^2  - bd} \right)g' - \alpha b^4 g''' + \frac{1}{3}\beta b^4 g'^3  + \frac{1}{2}\left( {\gamma b^2 d - b^2 d} \right)g'^2  + C_1 = 0 ,
\end{equation}
where $D$ is an arbitrary constant.

Letting $k_1  = \lambda d^2  + \delta b^2  - bd,k_2  = \frac{1}{3}\beta b^4 ,k_3  = \frac{1}{2}\left( {\gamma b^2 d - b^2 d} \right),k_4  = \alpha b^4$ , Eq.(\ref{ch-17}) can be reads as
\begin{equation}\label{ch-18}
k_1 g' + k_2 g'^3  + k_3 g'^2  - k_4 g''' + C_1  = 0 ,
\end{equation}

Setting $g'\left( \xi  \right) = p$ , Eq.(\ref{ch-18}) can be further simplified to
\begin{equation}\label{ch-19}
k_1 p + k_2 p^3  + k_3 p^2  - k_4 p'' + C_1  = 0 ,
\end{equation}

In order to get the solutions of Eq.(\ref{ch-19}) , We adopt two methods. The generalized tanh function expansion method\cite{24} to be adopted in first one . There are explicit solutions to the following form
\begin{equation}\label{ch-20}
p = \sum\limits_{i = 0}^n {q_i Y\left( \theta  \right)} ^i  + \sum\limits_{i = 1}^n {q_{ - i} Y\left( \theta  \right)} ^{ - i} ,
\end{equation}
where $\theta  = b\xi  - d\eta$ , and $n,q_i ,q_{ - i}$ are undetermined constants, also $Y$ is the solution of the $Riccati$ equation as follows
\begin{equation}\label{ch-21}
Y'\left( \theta  \right) = A + BY\left( \theta  \right) + CY^2 \left( \theta  \right) ,
\end{equation}
where $A,B$ and $C$ are constants.

Balancing the highest order derivative term and nonlinear term in Eq.(\ref{ch-19}) , it follows that $n = 1$. Therefore the explicit solution of the equation can be expressed as
\begin{equation}\label{ch-22}
p = q_0  + q_1 Y\left( \theta  \right) + \frac{{q_{ - 1} }}{{Y\left( \theta  \right)}},
\end{equation}

By substituting Eq.(\ref{ch-21}) and Eq.(\ref{ch-22}) into Eq.(\ref{ch-19}) , we can obtain the following solutions
\begin{equation}\label{ch-23}
\begin{array}{l}
p_1  = R + \sqrt {\frac{{2k_4 }}{{k_2 }}} C\left[ {\sec \left( \theta  \right) - \tan \left( \theta  \right)} \right] + \sqrt {\frac{{2k_4 }}{{k_2 }}} A\left[ {\sec \left( \theta  \right) - \tan \left( \theta  \right)} \right]^{ - 1} ,\\
p_2  = R + \sqrt {\frac{{2k_4 }}{{k_2 }}} C\frac{{\tan \theta }}{{1 \pm \sec \theta }} + \sqrt {\frac{{2k_4 }}{{k_2 }}} A\left( {\frac{{\tan \theta }}{{1 \pm \sec \theta }}} \right)^{ - 1} , \\
p_3  = R + \sqrt {\frac{{2k_4 }}{{k_2 }}} C\frac{{\exp (B\theta ) - A}}{B} + \sqrt {\frac{{2k_4 }}{{k_2 }}} A\left[ {\frac{{\exp (B\theta ) - A}}{B}} \right]^{ - 1} , \\
 \end{array}
\end{equation}
where $\theta  = b\xi  - d\eta$ and $R = \frac{{ - k_3  + \sqrt {k_3^2  - 12k_2 k_4 AC - 3k_1 k_2  + 3k_2 k_4 B^2 } }}{{3k_2 }}$ .

The solutions of the original equation are obtained, respectively,
\begin{equation}\label{ch-24}
\begin{array}{l}
u_1  = \int {F_2 \left( t \right)dt} + R\theta  + \sqrt {\frac{{2k_4 }}{{k_2 }}} C\left[ {\ln \left( {\sec \left( \theta  \right) + \tan \left( \theta  \right)} \right) + \ln \left( {\cos \left( \theta  \right)} \right)} \right] - \sqrt {\frac{{2k_4 }}{{k_2 }}} A\ln \left( {\sin \left( \theta  \right) - 1} \right) + C_2 ,\\
u_2  = \int {F_2 \left( t \right)dt} + R\theta  - \sqrt {\frac{{2k_4 }}{{k_2 }}} C\left[ {\ln \left( {\sec \left( \theta  \right) \pm 1} \right) - \ln \left( {\sec \left( \theta  \right)} \right)} \right] + \sqrt {\frac{{2k_4 }}{{k_2 }}} A\left[ {\ln \left( {\sin \left( \theta  \right)} \right) \pm \ln \left( {\csc \left( \theta  \right) - \cot \left( \theta  \right)} \right)} \right] + C_3  , \\
u_3  = \int {F_2 \left( t \right)dt} + R\theta  - \sqrt {\frac{{2k_4 }}{{k_2 }}} C\frac{{\frac{{e^{\left( {B\theta } \right)} }}{B} - A\theta }}{B} - \sqrt {\frac{{2k_4 }}{{k_2 }}} A\left[ {\frac{{\ln \left( { - e^{\left( {B\theta } \right)}  + A} \right)}}{A} - \frac{{\ln \left( {e^{\left( {B\theta } \right)} } \right)}}{A}} \right] + C_4 , \\
 \end{array}
\end{equation}
where $C_1 ,C_2 ,C_3$ are integral constants. When $\delta  = 0$, we get the Painleve  ${\rm I}{\rm I}$ solutions, but we get the other new exact solutions from this article.


The second, by applying Jacobi-elliptic function expansion method, we look for the solitary wave solutions of Eq.(\ref{ch-19}). Suppose the solutions has the following form
\begin{equation}\label{ch-25}
G = p\left( \theta  \right) = \sum\limits_{i = 0}^n {a_i } sn^i \left( \theta ,m \right) + \sum\limits_{i = 0}^n {b_i } sn^{ - i} \left( \theta ,m \right) .
\end{equation}

Balancing the highest order derivative term and nonlinear term in Eq.(\ref{ch-19}), it follows that $n = 1$ . Therefore, the explicit solution of Eq.(\ref{ch-25}) can be expressed as
\begin{equation}\label{ch-26}
p\left( \theta  \right) = a_0  + a_1 sn\left( \theta ,m \right) + b_1 sn^{ - 1} \left( \theta ,m \right) ,
\end{equation}
where $a_0 ,a_1 ,b_1$ are undetermined constants. In the same way, we can obtain the following solutions
\begin{equation}\label{ch-27}
\begin{array}{l}
G_1  = p_4  = \frac{{ - k_3  + \sqrt {k_3^2  - 18k_2 k_4 m - 3k_1 k_2  - 3k_2 k_4  - 3k_2 k_4 m^2 } }}{{3k_2 }} - \sqrt {\frac{{2k_4 }}{{k_2 }}} \left[ {sn\left( \theta ,m \right) - sn^{ - 1} \left( \theta ,m \right)} \right] ,\\
G_2  = p_5  = \frac{{ - k_3  + \sqrt {k_3^2  - 3k_2 k_4  - 3k_1 k_2  - 3k_2 k_4 m^2 } }}{{3k_2 }} - \sqrt {\frac{{2k_4 }}{{k_2 }}} msn\left( \theta ,m \right) , \\
 \end{array}
\end{equation}
where $0 < m < 1$ , $m$ is modulus of elliptic functions.

So we can get the other set of solutions for Eq.(\ref{ch-0})
\begin{equation}\label{ch-28}
\begin{array}{l}
u_4  = \int {F_2 \left( t \right)dt} + \frac{{ - k_3  + \sqrt {k_3^2  - 18k_2 k_4 m - 3k_1 k_2  - 3k_2 k_4  - 3k_2 k_4 m^2 } }}{{3k_2 }}\theta  - \sqrt {\frac{{2k_4 }}{{k_2 }}} \left[ {\frac{1}{m}\ln \left( {dn\left( \theta ,m \right)  - m cn\left( \theta ,m \right) } \right) - \ln \left( {\frac{{sn\left( \theta ,m \right) }}{{cn\left( \theta ,m \right)  + dn\left( \theta ,m \right) }}} \right)} \right] ,\\
u_5  = \int {F_2 \left( t \right)dt} + \frac{{ - k_3  + \sqrt {k_3^2  - 3k_2 k_4  - 3k_1 k_2  - 3k_2 k_4 m^2 } }}{{3k_2 }}\theta  - \sqrt {\frac{{2k_4 }}{{k_2 }}} \ln \left( {dn\left( {\theta ,m} \right) - mcn\left( {\theta ,m} \right)} \right) . \\
 \end{array}
\end{equation}
%

\textbf{Case 1.2}~~~~$a = 0,d \ne 0,b \ne 0,c \ne 0$ \\

In this case, we can get the following solution of Eq.(\ref{ch-12}) ,
\begin{equation}\label{ch-29}
\omega = \frac{{c\eta }}{b},\theta  = \xi , ~~~~~~~~f = \omega + h\left( \theta  \right) , \\
\end{equation}

Take Eq.(\ref{ch-29}) into Eq.(\ref{ch-12}), we can get
\begin{equation}\label{ch-30}
- \alpha h^{\left( 4 \right)}  + \beta h'^2 h'' + \left( {\frac{c}{b} + \delta } \right)h'' = 0 ,
\end{equation}

Setting $h'\left( \xi  \right) = W$ , and Eq.(\ref{ch-30})  can be written as
\begin{equation}\label{ch-31}
 - \alpha W'' + \frac{\beta }{3}W^3  + \left( {\frac{c}{b} + \delta } \right)W + C_5  = 0 ,
\end{equation}

Letting $A = \frac{\beta }{{3\alpha }},B = \frac{c}{{b\alpha }} + \frac{\delta }{\alpha },r = \frac{{C_5 }}{\alpha }$, then Eq.(\ref{ch-31}) can be written as
\begin{equation}\label{ch-32}
W'' = AW^3  + BW + R ,
\end{equation}

Taking different values of $A$ and $B$ , and Eq.(\ref{ch-32}) also have the corresponding different solutions,
 \begin{equation}\label{ch-33}
\begin{array}{l}
\left( 1 \right) If  A < 0,B > 0,R = 0, V_1  = \sqrt {\frac{{ - B}}{A}} \sec h\left( {\sqrt B \theta } \right) , \\
\left( 2 \right) If  A > 0,B < 0,R = 0, V_2  = \sqrt {\frac{{ - B}}{A}} \sec h\left( {\sqrt { - B} \theta } \right) , \\
\left( 3 \right) If  A > 0,B > 0,R = \frac{{B^2 }}{{4A}}, V_3  = \sqrt {\frac{B}{{2A}}} \sec h\left( {\sqrt {B/2} \theta } \right) , \\
\left( 4 \right) If  A = \frac{4}{m},B =  - 6m - m^2  - 1,R = 2m^3  + m^4  + m^2 , V_4  = \frac{{mdn\left( \theta ,m  \right)cn\left( \theta ,m  \right)}}{{msn^2 \left( \theta ,m \right) - 1}} ,  \\
\left( 5 \right) If  A =  - 4m_1 ,B =  - 6m_1  - m^2  + 2,R = 2 - 2m_1  - m^2 , V_5  =  - \frac{{m^2 sn\left( \theta ,m  \right)cn\left( \theta ,m \right)}}{{m_1  + dn^2 \left( \theta ,m \right)}} , \\
\left( 6 \right) If  A = m^2 \left( { - 1 + m^2 } \right),B = 2m^2  - 1,R = 1 , V_6  = sd\left( \theta  \right) = \frac{{sn\left( \theta ,m \right)}}{{dn\left( \theta ,m \right)}} ,  \\
\left( 7 \right) If  A = 1 - m^2 ,B = 2m^2  - 1,R =  - m^2, V_7  = nc\left( \theta ,m \right) = \frac{1}{{cn\left( \theta ,m \right)}} ,\\
\left( 8 \right) If  A = 1,B = 2 - m^2 ,R = 1 , V_8  = \frac{{sn\left( \theta ,m \right)dn\left( \theta ,m \right)}}{{cn\left( \theta ,m \right)}} , \\
\left( 9 \right) If  A = \frac{1}{4},B = \frac{{m^2  - 2}}{2},R = \frac{{m^2 }}{4} , V_9  = ns\left( \theta ,m \right) + ds\left( \theta ,m \right) , \\
\left( 10 \right) If  A = 1,B = 2 + m^2 ,R =  - 2m^2  + m^4  + 1 , V_{10}  = \frac{{dn\left( \theta ,m \right)cn\left( \theta ,m \right)}}{{sn\left( \theta ,m \right)}} , \\
 \end{array}
\end{equation}
where $m$ is modulus of elliptic functions.

Applying the above expressions of explicit solutions $W_i \left( {i = 1, \ldots 10} \right)$ , one can obtain $h$ and $f = \omega + h\left( \theta  \right)$ , and we can derive the corresponding solutions of Eq.(\ref{ch-0}). For example
 \begin{equation}\label{ch-343}
\begin{array}{l}
u_6  = \frac{{c\eta }}{b} + \sqrt {\frac{{ - 1}}{A}} \arctan \left( {\sinh \left( {\sqrt B \theta } \right)} \right) ,\\
u_7  = \frac{{c\eta }}{b} - \ln \left( {sn\left( {\theta ,m} \right)} \right) ,\\
u_8  = \frac{{c\eta }}{b} + \ln \left( {ds\left( {\theta ,m} \right) - cs\left( {\theta ,m} \right)} \right) + \ln \left( {ns\left( {\theta ,m} \right) - cs\left( {\theta ,m} \right)} \right). \\
 \end{array}
\end{equation}

\textbf{Case 2}~~~~Letting $c_2  = 1,c_1  = c_3  = c_4  = c_5  = 0,F_1 \left( t \right) = 0,F_2 \left( t \right) \ne 0$ , in the same way, we can get

\begin{equation}\label{ch-34}
\begin{array}{l}
\xi  = x,\eta  = y  ,
~~~~~~u = \int {F_2 \left( t \right)dt}  + f\left( {\xi ,\eta } \right) ,
 \end{array}
\end{equation}

Substituting Eq.(\ref{ch-34}) into Eq.(\ref{ch-0}) , we have
\begin{equation}\label{ch-35}
- \alpha f_{\xi \xi \xi \xi }  + \beta f_\xi  ^2 f_{\xi \xi }  - f_{\xi \xi } f_\eta   + \gamma f_\xi  f_{\xi \eta }  + \lambda f_{\eta \eta }  + \delta f_{\xi \xi }  = 0 .
\end{equation}

Making transformation with $f\left( {\xi ,\eta } \right) = f\left( w  \right),w  = l\xi  + k\eta$ , we can get
\begin{equation}\label{ch-36}
 - \alpha l^4 f^{\left( 4 \right)}  + \beta l^4 f'^2 f'' - l^2 kf'f'' + \gamma l^2 kf'f'' + \lambda k^2 f'' + \delta l^2 f'' = 0 .
\end{equation}

Integrating twice with respect to $f$ , yields
\begin{equation}\label{ch-37}
f''^2  = \frac{\beta }{{6\alpha }}f'^4  + \frac{{\left( {\gamma  - 1} \right)k}}{{\alpha l^2 }}f'^3  + \frac{{\lambda k^2  + \delta l^2 }}{{\alpha l^4 }}f'^2  + \frac{{2C_6 }}{{\alpha l^4 }}f' + \frac{{2C_7 }}{{\alpha l^4 }} = 0 ,
\end{equation}
where $d_0 ,d_1$ are arbitrary constants. Letting $f' = H$ , and Eq.(\ref{ch-37}) can be written as
\begin{equation}\label{ch-38}
H' = \mu \sqrt {h_0  + h_1 H + h_2 H^2  + h_3 H^3  + h_4 H^4 } ,
\end{equation}
where $f = f\left( w  \right),\mu  =  \pm 1$ , and
\begin{equation}\label{ch-39}
h_0  = \frac{{2C_7 }}{{\alpha l^4 }},h_1  = \frac{{2C_6 }}{{\alpha l^4 }},h_2  = \frac{{\lambda k^2  + \delta l^2 }}{{\alpha l^4 }},h_3  = \frac{{\left( {\gamma  - 1} \right)k}}{{\alpha l^2 }},h_4  = \frac{\beta }{{6\alpha }} .
\end{equation}

We can get the following solutions through the following discussion. The explicit solution of Eq.(\ref{ch-38}) have been given in \cite{22,23}.

\textbf{Case 2.1}~~Polynomial solutions

If $h_0  > 0,h_1  = h_2  = h_3  = h_4  = 0$, then $H = \mu \sqrt {h_0 } w  + C_8$, and we obtain
\begin{equation}\label{ch-40}
u_9  = \int {F_2 \left( t \right)dt} + \frac{1}{2}\mu \sqrt {h_0 } w ^2  + C_8 w.
\end{equation}

If $h_1  \ne 0,h_2  = h_3  = h_4  = 0$, then $H =  - \frac{{h_0 }}{{h_1 }} + \mu ^2 \left( {\frac{{h_1 }}{4}} \right)w ^2  + C_9$, then the solution of Eq.(\ref{ch-0}) is expressed by
\begin{equation}\label{ch-41}
u_{10}  = \int {F_2 \left( t \right)dt}  + \frac{1}{3}\mu ^2 \frac{{h_1 }}{4} w ^3  + \left( {C_9  - \frac{{h_0 }}{{h_1 }}} \right)w .
\end{equation}
where $w  = l\xi  + k\eta$, and $\xi, \eta$ are arbitrary constants.

\textbf{Case 2.2}~~Rational solutions

If $h_4  > 0,h_0  = h_1  = h_2  = h_3  = 0$, then $H =  - \frac{1}{{\mu \sqrt {h_4 } w }} + C_{10}$, then we obtain
\begin{equation}\label{ch-42}
u_{11}  = \int {F_2 \left( t \right)dt}  + \frac{{\ln w }}{{\mu \sqrt {h_4 } }} + C_{10} w .
\end{equation}

If $h_2  > 0,h_0  = h_1  = h_3  = h_4  = 0$, then $H = e^{\mu \sqrt {h_2 w } }  + C_{11}$, and one can derive
\begin{equation}\label{ch-43}
u_{12}  = \int {F_2 \left( t \right)dt}  + \frac{{2\sqrt w  e^{\mu \sqrt {h_2 w } } }}{{\mu \sqrt {h_2 } }} + C_{11} w .
\end{equation}
where $w  = l\xi  + k\eta$, and $\xi, \eta$ are arbitrary constants.

\textbf{Case 2.3}~~Triangular function solutions

$\left( 1 \right)$  If $h_0  = h_1  = 0$, then we can get
\begin{equation}\label{ch-44}
\begin{array}{l}
 H_1 \left( w  \right) = \frac{{4h_2 C_0 \sec h^2 \left( {\frac{{\sqrt {h_2 } }}{2}w } \right)}}{{2\mu \left( {1 - \Delta _1 } \right) + 2\mu \left( {1 + \Delta _1 } \right)\tanh \left( {\frac{{\sqrt {h_2 } }}{2}w } \right) - \left( {2h_3 C_0  + 1 - \Delta _1 } \right)\sec h^2 \left( {\frac{{\sqrt {h_2 } }}{2}w } \right)}} , \\
H_2 \left( w  \right) = \frac{{4h_2 C_0 \sec h^2 \left( {\frac{{\sqrt {h_2 } }}{2}w } \right)}}{{2\mu \Theta  - 2\mu \left( {\Delta _2  + C_0^2 } \right)\tanh \left( {\frac{{\sqrt {h_2 } }}{2}w } \right) - \left( {2h_3 C_0  + \Theta } \right)\sec h^2 \left( {\frac{{\sqrt {h_2 } }}{2}w } \right)}}, \\
 \end{array}
\end{equation}
where the constants $h_2  > 0$ , $C_0  = \exp \left( {\sqrt {h_2 } C_{12} } \right),\Delta _1  = C_0^2 \left( {4h_2 h_4  - h_3^2 } \right),\Delta _2  = 4h_2 h_4  - h_3^2$ and $\Theta  = C_0^2  - \Delta _2$.

Then the solution of Eq.(\ref{ch-0}) is expressed by
\begin{equation}\label{ch-45}
\begin{array}{l}
u_{13}  = \int {F_2 \left( t \right)dt}  + \int {\frac{{4h_2 C_0 \sec h^2 \left( {\frac{{\sqrt {h_2 } }}{2}w } \right)}}{{2\mu \left( {1 - \Delta _1 } \right) + 2\mu \left( {1 + \Delta _1 } \right)\tanh \left( {\frac{{\sqrt {h_2 } }}{2}w } \right) - \left( {2h_3 C_0  + 1 - \Delta _1 } \right)\sec h^2 \left( {\frac{{\sqrt {h_2 } }}{2}w } \right)}}} dw  + C_{13} , \\
u_{14}  = \int {F_2 \left( t \right)dt}  + \int {\frac{{4h_2 C_0 \sec h^2 \left( {\frac{{\sqrt {h_2 } }}{2}w } \right)}}{{2\mu \Theta  - 2\mu \left( {\Delta _2  + C_0^2 } \right)\tanh \left( {\frac{{\sqrt {h_2 } }}{2}w } \right) - \left( {2h_3 C_0  + \Theta } \right)\sec h^2 \left( {\frac{{\sqrt {h_2 } }}{2}w } \right)}}} dw  + C_{14} . \\
 \end{array}
\end{equation}

$\left( 2 \right)$  If $h_1  = h_3  = 0,h_0  = \frac{{h_2^2 }}{{4h_4 }},h_2  > 0,h_4  > 0$, then $H = \mu \sqrt {\frac{{h_2 }}{{h_4 }}} \tan \left( {\sqrt {\frac{{h_2 }}{2}} w } \right)$ , we can get
\begin{equation}\label{ch-46}
u_{15}  = \int {F_2 \left( t \right)dt} + \mu \sqrt {\frac{1}{{2h_4 }}} \ln \left( {1 + \tan ^2 \left( {\sqrt {\frac{{h_2 }}{2}} w } \right)} \right) + C_{15} . \\
\end{equation}
where $w  = l\xi  + k\eta$, and $\xi, \eta$ are arbitrary constants.

In order to reflect the characteristics and properties of the solution $u_{15}$, we give the image of the solution $u_{15}$, as follows


\textbf{Case 2.4}~~Elliptic periodic solutions

If $h_1  = h_3  = 0$ , then we can obtain some new Jacobi-elliptic function solutions

$\left( 1 \right)$
\begin{equation}\label{ch-47}
H_3  = \mu \sqrt {\frac{{ - h_2 m^2 }}{{h_4 \left( {2m^2  - 1} \right)}}} cn\left( {\sqrt {\frac{{h_2 }}{{2m^2  - 1}}} w } ,m \right),h_4  < 0,h_2  > 0,h_0  = \frac{{h_2^2 m^2 \left( {1 - m^2 } \right)}}{{h_4 \left( {2m^2  - 1} \right)^2 }} .
\end{equation}

So Eq.(\ref{ch-0}) has the following solution
\begin{equation}\label{ch-48}
u_{16}  = \int {F_2 \left( t \right)dt} + \mu \sqrt {\frac{{ - h_2 m^2 }}{{h_4 \left( {2m^2  - 1} \right)}}} \int {cn\left( {\sqrt {\frac{{h_2 }}{{2m^2  - 1}}} w } ,m \right)dw  + C_{16} }  .
\end{equation}

$\left( 2 \right)$
\begin{equation}\label{ch-49}
H_4 \left( w  \right) = \mu \sqrt {\frac{{ - h_2 }}{{h_4 \left( {2 - m^2 } \right)}}} dn\left( {\sqrt {\frac{{h_2 }}{{2 - m^2 }}} w  } ,m \right),h_4  < 0,h_2  > 0,h_0  = \frac{{h_2^2 \left( {1 - m^2 } \right)}}{{h_4 \left( {2 - m^2 } \right)^2 }} ,
\end{equation}

And one can derive
\begin{equation}\label{ch-50}
u_{17}  = \int {F_2 \left( t \right)dt} + \mu \sqrt {\frac{{ - h_2 }}{{h_4 \left( {2 - m^2 } \right)}}} \int {dn\left( {\sqrt {\frac{{h_2 }}{{2 - m^2 }}} w  } ,m \right)dw }  + C_{17}  .
\end{equation}

$\left( 3 \right)$
\begin{equation}\label{ch-51}
H_5 \left( w  \right) = \mu \sqrt {\frac{{ - h_2 m^2 }}{{h_4 \left( {m^2  + 1} \right)}}} sn\left( {\sqrt { - \frac{{h_2 }}{{m^2  + 1}}} w } , m \right),h_4  > 0,h_2  < 0,h_0  = \frac{{h_2^2 m^2 }}{{h_4 \left( {m^2  + 1} \right)^2 }} ,
\end{equation}

And we can obtain
\begin{equation}\label{ch-52}
u_{18}  = \int {F_2 \left( t \right)dt} + \mu \sqrt {\frac{{ - h_2 m^2 }}{{h_4 \left( {m^2  + 1} \right)}}} \int {sn\left( {\sqrt { - \frac{{h_2 }}{{m^2  + 1}}} w } , m \right)dw }  + C_{18}  .
\end{equation}
where $0 < m < 1$, $m$ is modulus of elliptic functions, $w  = l\xi  + k\eta$, and $\xi, \eta$ are arbitrary constants.

When $m \to 1$ and $m \to 0$ㄛThe Jacobi elliptic function is degenerated into trigonometric or hyperbolic functionㄛas follows

$m \to 1,sn\left( {w,m} \right) \to \tanh \left( w \right),cn\left( {w,m} \right) \to \sec h\left( w \right),dn\left( {w,m} \right) \to \sec h\left( w \right),$

$m \to 0,sn\left( {w,m} \right) \to \sin \left( w \right),cn\left( {w,m} \right) \to \cos \left( w \right),dn\left( {w,m} \right) \to 1$.

\textbf{Case 2.5}~~Weierstrass periodic solutions

If $h_2  = h_4  = 0$, then Eq.(\ref{ch-38}) can get some Weierstrass periodic solutions

$H_6 \left( w  \right) = \mu \wp \left( {\frac{{\sqrt {h_3 } }}{2}w ,g_2 ,g_3 } \right)$, where $h_3  > 0,g_2  =  - 4\frac{{h_1 }}{{h_3 }},g_3  =  - 4\frac{{h_0 }}{{h_3 }}$ .

So we can derive solution of Eq.(\ref{ch-0})
\begin{equation}\label{ch-53}
u_{19}  = \int {F_2 \left( t \right)dt} + \mu \int {\wp \left( {\frac{{\sqrt {h_3 } }}{2}w ,g_2 ,g_3 } \right)dw }  + C_{19} .
\end{equation}

\textbf{Case 3}~~~~Letting $c_2  = c_4  = 1,c_1  = c_3  = c_5  = 0,F_1 \left( t \right) = a,F_2 \left( t \right) = b$ , solving the corresponding characteristic equations, we have the following invariants $\xi,\eta$ and the function $u$ .
\begin{equation}\label{ch-54}
\begin{array}{l}
\xi  = y + \frac{x}{a},\eta  = t + \frac{x}{a}  ,
~~~~~~u =  - \frac{b}{a}x + f\left( {\xi ,\eta } \right) ,
 \end{array}
\end{equation}

Substituting Eq.(\ref{ch-54}) into Eq.(\ref{ch-0}) , we have
\begin{equation}\label{ch-55}
\begin{array}{l}
 \left( {\frac{{\lambda a^2  + \delta  - b\gamma }}{{a^2 }} + \frac{{b^2 \beta }}{{a^4 }}} \right)f_{\xi \xi }  + \left( {\frac{{2a - b\gamma  + 2\delta }}{{a^2 }} + \frac{{2b^2 \beta }}{{a^4 }}} \right)f_{\xi \eta }  + \left( {\frac{{a + \delta }}{{a^2 }} + \frac{{b^2 \beta }}{{a^4 }}} \right)f_{\eta \eta }  + \left( {\frac{{\gamma  - 1}}{{a^2 }} - \frac{{2b\beta }}{{a^4 }}} \right)f_\xi  f_{\xi \xi }  \\
  + \left( {\frac{{\gamma  - 2}}{{a^2 }} - \frac{{4b\beta }}{{a^4 }}} \right)f_\xi  f_{\xi \eta }  - \left( {\frac{1}{{a^2 }} + \frac{{2b\beta }}{{a^4 }}} \right)f_\xi  f_{\eta \eta }  + \left( {\frac{\gamma }{{a^2 }} - \frac{{4b\beta }}{{a^4 }}} \right)f_\eta  f_{\xi \eta }  + \left( {\frac{\gamma }{{a^2 }} - \frac{{2b\beta }}{{a^4 }}} \right)f_\eta  f_{\xi \xi }  - \frac{{4b\beta }}{{a^4 }}f_\eta  f_{\eta \eta }  \\
  - \frac{\alpha }{{a^4 }}\left( {f_{\xi \xi \xi \xi }  + 4f_{\xi \xi \xi \eta }  + 6f_{\xi \xi \eta \eta }  + 4f_{\xi \eta \eta \eta }  + f_{\eta \eta \eta \eta } } \right) + \frac{\beta }{{a^4 }}\left( {f_\xi   + f_\eta  } \right)^2 \left( {f_{\xi \xi }  + f_{\eta \eta }  + 2f_{\xi \eta } } \right) = 0  . \\
 \end{array}
\end{equation}

Making transformation with $f\left( {\xi ,\eta } \right) = f\left( \Omega  \right),\Omega  = k\xi  + l\eta$ ,we can get
\begin{equation}\label{ch-56}
\begin{array}{l}
 \left( {\frac{{kl\left( {2a - b\gamma  + 2\delta } \right) + \left( {a + \delta } \right)l^2  + \left( {\lambda a^2  + \delta  - b\gamma } \right)k^2 }}{{a^2 }} + \frac{{b^2 \beta \left( {l + k} \right)^2 }}{{a^4 }}} \right)f''
  + \left( {\frac{{\left( {\gamma  - 1} \right)\left( {k^3  + 2lk^2  + l^2 k} \right)}}{{a^2 }} - \frac{{2b\beta \left( {k + l} \right)^3 }}{{a^4 }}} \right)f'f'' \\
  + \frac{{\beta \left( {l + k} \right)^4 }}{{a^4 }}f'^2 f'' - \frac{{\alpha \left( {l + k} \right)^4 }}{{a^4 }}f^{\left( 4 \right)}  = 0 .\\
 \end{array}
\end{equation}

Integrating twice with respect to $f$ ,yields
\begin{equation}\label{ch-57}
\begin{array}{l}
 f''^2  = \frac{\beta }{{6\alpha }}f'^4  + \left( {\frac{{a^2 \left( {\gamma  - 1} \right)\left( {k^3  + 2lk^2  + l^2 k} \right)}}{{3\alpha (l + k)^4 }} - \frac{{2b\beta }}{{3\alpha \left( {l + k} \right)}}} \right)f'^3  \\
  ~~~~~~~~~~+ \left( {\frac{{a^2 \left[ {kl\left( {2a - b\gamma  + 2\delta } \right) + \left( {a + \delta } \right)l^2  + \left( {\lambda a^2  + \delta  - b\gamma } \right)k^2 } \right]}}{{\alpha \left( {l + k} \right)^4 }} + \frac{{2b^2 \beta }}{{\alpha \left( {l + k} \right)^2 }}} \right)f'^2  \\
  ~~~~~~~~~~+ \frac{{2a^4 e_0 }}{{\alpha \left( {l + k} \right)^4 }}f' + \frac{{2a^4 e_1 }}{{\alpha \left( {l + k} \right)^4 }}  . \\
 \end{array}
\end{equation}

Eq.(\ref{ch-37}) and Eq.(\ref{ch-57}) are all elliptic functions, so the solutions of case 3 is reference to case 2 .

\textbf{Case 4}~~~~Letting $c_4  = 1,c_1  = c_2  = c_3  = c_5  = 0,F_1 \left( t \right) = 0,F_2 \left( t \right) = t^2$ , in the same way, one can derive
\begin{equation}\label{ch-58}
\begin{array}{l}
\xi  =  x, \eta  =  t  ,
~~~~~~u = t^2 y + f\left( {\xi ,\eta } \right) ,
 \end{array}
\end{equation}

Substituting Eq.(\ref{ch-58}) into Eq.(\ref{ch-0}), we get
\begin{equation}\label{ch-59}
f_{\xi \eta }  - \alpha f_{\xi \xi \xi \xi }  + \beta f_\xi  ^2 f_{\xi \xi }  - \eta ^2 f_{\xi \xi }  + \delta f_{\xi \xi }  = 0 .
\end{equation}

Assuming Eq.(\ref{ch-59}) has the following solution
\begin{equation}\label{ch-60}
f = s\left( \eta  \right)\xi  + z\left( \eta  \right) ,
\end{equation}
where $s\left( \eta  \right)$ and $z\left( \eta  \right)$ are the functions to be determined.  Substituting Eq.(\ref{ch-60}) into Eq.(\ref{ch-59}) yields
\begin{equation}\label{ch-61}
f = Z_1 \xi  + z\left( t \right) .
\end{equation}

Combining Eq.(\ref{ch-58}) and Eq.(\ref{ch-61}) , then the new exact solution of the Eq.(\ref{ch-0})
\begin{equation}\label{ch-62}
u_{20}  = t^2 y + Z_1 x + z\left( t \right) .
\end{equation}
where $Z_1$ is integral constant.

\section{Conservation laws}\label{Conservation laws}

In this section, we will study the conservation laws by using the adjoint equation and symmetries of Eq.(\ref{ch-0}). For Eq.(\ref{ch-0}), the adjoint equation has the form
\begin{equation}\label{ch-63}
v_{xt}  - \alpha v_{xxxx}  + \beta u_x^2 v_{xx}  - 2\beta u_{xx} u_x v_x  - v_{xx} u_y  + u_{xx} v_y  - \gamma v_x u_{xy}  + \gamma u_x v_{xy}  + \lambda v_{yy}  + \delta v_{xx}  = 0 .
\end{equation}
and the Lagrangian in the symmetrized form
\begin{equation}\label{ch-64}
L = v\left( {u_{xt}  - \alpha u_{xxxx}  + \beta u_x^2 u_{xx}  - u_{xx} u_y  + \gamma u_x u_{xy}  + \lambda u_{yy}  + \delta u_{xx} } \right) .
\end{equation}
\textbf{Theorem 1.} Every Lie point, Lie-B$\ddot{a}$clund and non-local symmetry of  Eq.(\ref{ch-0}) provides a conservation law for  Eq.(\ref{ch-0}) and the adjoint equation. Then the elements of conservation vector $(C^1,C^2,C^3 )$ are defined by the following expression:
\begin{equation}\label{ch-65}
\begin{array}{l}
 C^i  = \xi ^i L + W^\alpha  \left[ {\frac{{\partial L}}{{\partial u_{ij}^\alpha  }} - D_j \left( {\frac{{\partial L}}{{\partial u_{ij}^\alpha  }}} \right) + D_j D_k \left( {\frac{{\partial L}}{{\partial u_{ijk}^\alpha  }}} \right) - D_j D_k D_r \left( {\frac{{\partial L}}{{\partial u_{ijkr}^\alpha  }}} \right) +  \cdots } \right] \\
  ~~~~~~~~+ D_j \left( W \right)\left[ {\frac{{\partial L}}{{\partial u_{ij}^\alpha  }} - D_k \left( {\frac{{\partial L}}{{\partial u_{ijk}^\alpha  }}} \right) +  \cdots } \right] \\
  ~~~~~~~~+ D_j D_k \left( W \right)\left[ {\frac{{\partial L}}{{\partial u_{ijk}^\alpha  }} - D_r \left( {\frac{{\partial L}}{{\partial u_{ijkr}^\alpha  }}} \right) +  \cdots } \right] ,\\
 \end{array}
\end{equation}
where $W$ is the Lie characteristic function and
\begin{equation}\label{ch-66}
W^\alpha = \eta_\ast^{\alpha} - \xi^{j} u_j^{\alpha} .
\end{equation}

The conserved vector corresponding to an operator is
\begin{equation}\label{ch-67}
v = \xi ^1 \frac{\partial }{{\partial t}} + \xi ^2 \frac{\partial }{{\partial x}} + \xi ^3 \frac{\partial }{{\partial y}} + \xi ^4 \frac{\partial }{{\partial u}} .
\end{equation}
The operator $v$ yields the conservation laws $D_t \left( {C^1 } \right) + D_x \left( {C^2 } \right) + D_y \left( {C^3 } \right) = 0$, where the conserved vector $C = \left( {C^1 ,C^2 ,C^3 } \right)$ is given by Eq.(\ref{ch-65}) and has the components
\begin{equation}\label{ch-68}
\begin{array}{l}
 C^1  = \xi ^1 L + W\left( {\frac{{\partial L}}{{\partial u_t }} - D_x \left( {\frac{{\partial L}}{{\partial u_{xt} }}} \right)} \right) + W_x \left( {\frac{{\partial L}}{{\partial u_{xt} }}} \right), \\
 C^2  = \xi ^2 L + W\left( {\frac{{\partial L}}{{\partial u_x }} - D_x \left( {\frac{{\partial L}}{{\partial u_{xx} }}} \right) - D_y \left( {\frac{{\partial L}}{{\partial u_{xy} }}} \right) - D_t \left( {\frac{{\partial L}}{{\partial u_{xt} }}} \right) - D_{xxx} \left( {\frac{{\partial L}}{{\partial u_{xxxx} }}} \right)} \right) \\
  ~~~~~~~~+ W_x \left( {\frac{{\partial L}}{{\partial u_{xx} }} + D_{xx} \left( {\frac{{\partial L}}{{\partial u_{xxxx} }}} \right)} \right) + W_{xx} \left( { - D_x \left( {\frac{{\partial L}}{{\partial u_{xxxx} }}} \right)} \right) + W_{xxx} \left( {\frac{{\partial L}}{{\partial u_{xxxx} }}} \right) \\
 C^3  = \xi ^3 L + W\left( {\frac{{\partial L}}{{\partial u_y }} - D_x \left( {\frac{{\partial L}}{{\partial u_{xy} }}} \right) - D_y \left( {\frac{{\partial L}}{{\partial u_{yy} }}} \right)} \right) + W_x \left( {\frac{{\partial L}}{{\partial u_{xy} }}} \right) + W_y \left( {\frac{{\partial L}}{{\partial u_{yy} }}} \right) .\\
 \end{array}
\end{equation}
So Eq.(\ref{ch-68}) defined the corresponding components of non-local conservation law for the system of Eq.(\ref{ch-0}) and Eq.(\ref{ch-63}) corresponding to any operator admitted by Eq.(\ref{ch-0}).

Let us make conclusions for the operator
\begin{equation}\label{ch-69}
\begin{array}{l}
 V = \left( {\frac{{c_1 }}{3}x - F_1 \left( t \right) + \frac{{2c_1 \delta t}}{3} + \frac{{2\beta c_3 y}}{{ - 4\beta \lambda  + \gamma ^2  - \gamma }} + c_5 } \right)\frac{\partial }{{\partial x}} + \left( {c_1 t + c_2 } \right)\frac{\partial }{{\partial t}} + \left( {\frac{{2c_1 }}{3}y + c_3 t + c_4 } \right)\frac{\partial }{{\partial y}} \\ ~~~~~~~~+ \left( {\frac{{c_3 \left( {\gamma  - 1} \right)x}}{{ - 4\beta \lambda  + \gamma ^2  - \gamma }} + F_1 ^\prime  \left( t \right)y + F_2 \left( t \right)} \right)\frac{\partial }{{\partial u}} .\\
\end{array}
\end{equation}

For this operator, one can get
\begin{equation}\label{ch-70}
\begin{array}{l}
W = \frac{{\left( {\gamma  - 1} \right)x}}{{ - 4\beta \lambda  + \gamma ^2  - \gamma }} - \frac{{2\beta y}}{{ - 4\beta \lambda  + \gamma ^2  - \gamma }}u_x  - tu_y .\\
\end{array}
\end{equation}

We can get the conservation vector of Eq.(\ref{ch-0})
\begin{equation}\label{ch-71}
\begin{array}{l}
 C^1  =  - \left( {\frac{{\left( {\gamma  - 1} \right)x}}{{ - 4\beta \lambda  + \gamma ^2  - \gamma }} - \frac{{2\beta y}}{{ - 4\beta \lambda  + \gamma ^2  - \gamma }}u_x  - tu_y } \right)v_x  + \left( {\frac{{\left( {\gamma  - 1} \right)}}{{ - 4\beta \lambda  + \gamma ^2  - \gamma }} - \frac{{2\beta y}}{{ - 4\beta \lambda  + \gamma ^2  - \gamma }}u_{xx}  - tu_{xy} } \right)v , \\
 C^2  =  - \frac{{2\beta y}}{{ - 4\beta \lambda  + \gamma ^2  - \gamma }}v\left( {u_{xt}  - \alpha u_{xxxx}  + \beta u_x^2 u_{xx}  - u_{xx} u_y  + \gamma u_x u_{xy}  + \lambda u_{yy}  + \delta u_{xx} } \right) \\
  ~~~~~~~~+ \left( {\frac{{\left( {\gamma  - 1} \right)x}}{{ - 4\beta \lambda  + \gamma ^2  - \gamma }} - \frac{{2\beta y}}{{ - 4\beta \lambda  + \gamma ^2  - \gamma }}u_x  - tu_y } \right)\left( {u_{xy} v + \beta v_x u_x^2  - v_x u_y  + \delta v_x  - \gamma v_y u_x  - v_t  + \alpha v_{xxx} } \right) \\
  ~~~~~~~~+ \left( {\frac{{\left( {\gamma  - 1} \right)}}{{ - 4\beta \lambda  + \gamma ^2  - \gamma }} - \frac{{2\beta y}}{{ - 4\beta \lambda  + \gamma ^2  - \gamma }}u_{xx}  - tu_{xy} } \right)\left( {\beta vu_x^2  - vu_y  + v\delta  - \alpha v_{xx} } \right) \\
  ~~~~~~~~- \left( {\frac{{2\beta y}}{{ - 4\beta \lambda  + \gamma ^2  - \gamma }}u_{xxx}  + tu_{xxy} } \right)\alpha v_x  + \left( {\frac{{2\beta y}}{{ - 4\beta \lambda  + \gamma ^2  - \gamma }}u_{xxxx}  + tu_{xxxy} } \right)\alpha v  ,\\
 C^3  =  - tv\left( {u_{xt}  - \alpha u_{xxxx}  + \beta u_x^2 u_{xx}  - u_{xx} u_y  + \gamma u_x u_{xy}  + \lambda u_{yy}  + \delta u_{xx} } \right) \\
  ~~~~~~~~+ \left( {\frac{{\left( {\gamma  - 1} \right)x}}{{ - 4\beta \lambda  + \gamma ^2  - \gamma }} - \frac{{2\beta y}}{{ - 4\beta \lambda  + \gamma ^2  - \gamma }}u_x  - tu_y } \right)\left( { - vu_{xx}  - \gamma v_x u_x  - \gamma vu_{xx}  - \lambda v_y } \right) \\
  ~~~~~~~~+ \left( {\frac{{\left( {\gamma  - 1} \right)}}{{ - 4\beta \lambda  + \gamma ^2  - \gamma }} - \frac{{2\beta y}}{{ - 4\beta \lambda  + \gamma ^2  - \gamma }}u_{xx}  - tu_{xy} } \right)\gamma vu_x  - \left( {\frac{{2\beta }}{{ - 4\beta \lambda  + \gamma ^2  - \gamma }}u_x  + \frac{{2\beta y}}{{ - 4\beta \lambda  + \gamma ^2  - \gamma }}u_{xy}  + tu_{yy} } \right)\lambda v  .\\
 \end{array}
\end{equation}
This vector involves an arbitrary solution $v$ of the adjoint equation Eq.(\ref{ch-63}) and provides an infinite number of the conservation laws.

\textbf{Remark } We have checked all the exact solutions satisfy the original Eq.(\ref{ch-0}) , and the above solutions of the extended  (2+1)-dimensional Jaulent-Miodek equation have not been found in other references. To our knowledge, the symmetry reductions obtained in this paper are completely new and has not been studied yet in previous papers.

\section{conclusion}\label{conclusion}

In this paper, we investigated the extended  (2+1)-dimensional Jaulent-Miodek equation via Lie symmetries. By applying the direct symmetry method ,we obtain the classical Lie point symmetry of the equation. Based on given Lie symmetry , we reduce the extended  (2+1)-dimensional Jaulent-Miodek equation to the  (1+1)- dimensional PDE and ODE. We also have derived some exact solutions of Eq.(\ref{ch-0}) by using the relationship between the new solutions and the old ones. In order to reflect the characteristics and properties of this solutions, we give figures of some solutions. At last, we give the conservation laws of Eq.(\ref{ch-0}). These conclusions may be useful for the explanation of some practical physical problems.

The symmetry study plays a very important role in almost all scientific fields, especially in the soliton theory because of the existence of infinitely many symmetries for integrable system. Furthermore, the symmetry analysis based on the Lie group method is a very powerful method and is worthy of being studyied further.

\section{Acknowledgments}
This work is supported by National Natural Science Foundation of China under Grant(Nos.11505090), Research Award Foundation for Outstanding Young Scientists of Shandong Province(No.BS2015SF009) .

\begin{flushleft}
\textbf{References}
\end{flushleft}







\end{document}